Effects of vibration and rigidity modes of motion on the spectral statistics of spherical nuclei

H. Sabri[*], A. Hosseinnezhad

Department of Physics, University of Tabriz, Tabriz 51664, Iran.

[*] Corresponding author: h-sabri@tabrizu.ac.ir

**Abstract**

In this paper, we investigated the effects of *β*-vibration and *β*-rigidity on the energy levels from the viewpoint of statistical fluctuations of nuclear systems. To this aim, a parameter-free collective solution of the Bohr Hamiltonian in the five-dimensional harmonic oscillator potential with a linear energy dependence and an asymptotic limit of the slope are used to determine all of the observed normal states in even-even nuclei with $R_{4_1^+/2_1^+}$ ~ 2.00 - 2.15 ratio in the A ~ 90 -140 mass region. Different sequences are prepared of the energy levels, both experimental values and theoretical predictions, which are categorized as their spin-parity, *β* oscillator quanta, and seniority numbers and analyzed in the framework of random matrix theory to show their statistical situation in comparison with regular and correlated limits. Also, up to 2226 levels with the same $2^+$ spin-parity assignment are determined for different systems in which the stiffness parameter for them changed between *a* = 0 and *a* =1 limits and then analyzed in the same process. The results showed a transition between correlated behavior and regularity when the rigidity increased in considered systems. Also, there are apparent relations between the chaocity degrees of considered sequences and the considered criteria for classifications.

1. Introduction

The competition between the single particle and collective degrees of freedom is regarded as the main factor in the spectroscopic properties of medium- and heavy-mass nuclei. A combination of collective and single-particle motions in the different nuclei make their description so complicated. The nuclei which, are located in the vicinity of proton (or neutron) closed shells show spherical shape and therefore, the investigation about their structure and their level scheme makes obvious knowledge about the vibration and rigidity as the two main modes of their collective motion [1-5]. In this way, the Bohr-Mottelson collective model (BMM) [1], which consider the nuclear structure by using two geometrical variables, *β* & *γ*, and the interacting boson model (IBM) [3], which uses the algebraic structure in the framework of U(6) lie algebra, are regarded as the most powerful tools for such investigations. The spherical nuclei, which are the subject of our current study, correspond with the U(5) limit of IBM and equivalently, in the framework of BMM, these nuclei are described by the spherical vibrator model, which is based on a five-dimensional harmonic oscillator potential in the *β* variable. These nuclei are the subject of different theoretical and experimental studies in recent years, and tried to describe some phenomena such as long-lived isomeric states, the octupole shape coexistence and four quasiparticle (4qp) states observed in the context of them [5-19].

The Random matrix theory (RMT) is known as the most powerful tool in the statistical analyses of samples and has been widely used in different branches of physics, economics, and other subjects, which are explained in detail in various literatures such as [20-41]. RMT and its different measures, such as nearest neighbor spacing distribution (NNSD) and $\Delta_3(L)$ measure make it possible to get a correlation of the considered samples in comparison with three ensembles of this theory, Gaussian orthogonal, Gaussian unitary, and Gaussian symplectic ensembles [34-46]. On the other hand, the statistical situation of the systems are far from correlation, described via Poisson distribution. In recent years, the even-even nuclei have been the subject of different studies via RMT which present the correlation of the energy levels and electromagnetic transition rates in comparison with different parameters such as mass region, spin and deformation [23,34,36,38-39] which used the positive and negative parity states. These studies provided significant insights about nuclear structures, for example, spherical nuclei which have rotational symmetry, have the maximum correlation but deformed nuclei which have a mix of different types of symmetries explore Poisson statistics [46-50]. In this study, we tried to find a meaningful relation between the spectral statistics of different nuclear systems and two different modes in radial dependence of potential, vibration and rigidity similar to investigations presented in Refs.[42-44]. To this aim and to control possible relations independent of the particular hypothesis of theoretical predictions, the latest available empirical data [51] are used to prepare different sequences. Also, the parameter- free collective solution of the Bohr Hamiltonian which has a linear energy dependence in the five-dimensional harmonic oscillator potential and the asymptotic limit of the slope [5] is used to determine the energy levels of even-even nuclei which are classified as the U(5) candidate in the A ~ 90 -140 mass region. Different sequences are prepared by using these levels which are classified as their quantum numbers. Also, the $2^+$ levels which are calculated theoretically for such systems which have N = 25, 50 and 100 and assigned different stiffness parameters to them between $\beta$ vibration and $\beta$ rigidity limits are analyzed in the same manner to consider their statistical properties. The NNSD measure of RMT together with the Berry-Robnik distribution have been used to describe the chaocity measure of these sequences in comparison with both regular and correlated limits.

2. Energy levels and data sets

Three U(5), O(6) and SU(3) dynamical symmetry limits of IBM have been used to describe energy spectra of respectively, spherical, $\gamma$- unstable and deformed nuclei. A common measure for the classification of different nuclei in the framework of these symmetry limits is the ratio of the first $4^+$ and $2^+$ energy levels, $R_{4/2} \approx \dfrac{E(4_1^+)}{E(2_1^+)}$, which its values, 2.00, 2.50 and 3.33 are correspond with the above-mentioned symmetries, respectively [2-3]. Also, two 2.20 and 2.91 values of this ratio are used to identify such nuclei which are suggested as the candidates for E(5) and X(5) critical symmetries, respectively, the critical points of U(5)↔O(6) and U(5) ↔SU(3) transitional regions [6-7]. Based on these classifications and to consider such nuclei that have underlying U(5) symmetry and can be described by the spherical vibrator model, we focus on the experimental $0^+$, $2^+$ and $4^+$ energy levels of nuclei whose $R_{4/2}$ ratio are in the 2.00 ~ 2.15 region in the A ~90 -140 mass region and E ≤ 3 MeV. These regions are selected due to the majority of spherical nuclei in the vicinity of Z = 50 proton closed shell and also to consider energy levels far from the resonance region and can be analyzed by RMT. These nuclei which have at least five consecutive levels with definite spin-parity are, $^{90,94-100}$Mo, $^{90-98}$Ru, $^{92-94,98-100}$Pd, $^{98-102}$Cd, $^{110-116, 124-126}$Sn, $^{100-130}$Te, $^{132-140}$Xe, $^{138-140}$Ba and $^{140}$Ce. On the other hand and to determine these levels theoretically, we have followed the prescription introduced by Budaca for the energy levels of such nuclei. Details about this method are available in Ref.[5] and here, we briefly outline the basic ansatz and summarize the results. The original Bohr Hamiltonian is as follows:

$$\hat{H} = -\frac{\hbar^2}{2B}\left[\frac{1}{\beta^4}\frac{\partial}{\partial\beta}\beta^4\frac{\partial}{\partial\beta} + \frac{1}{\beta^2\sin 3\gamma}\frac{\partial}{\partial\gamma}\sin 3\gamma\frac{\partial}{\partial\gamma} - \frac{1}{4\beta^2}\sum_{k=1,2,3}\frac{Q_k^2}{\sin^2(\gamma - \frac{2}{3}\pi k)}\right] + V(\beta,\gamma) \quad , \tag{1}$$

In this notation, the usual collective coordinates are described by $\beta$ and $\gamma$, $Q_k$ ($k = 1, 2, 3$) presents the components of angular momentum and the mass parameter is $B$. For nuclei whose potentials are independent of $\gamma$ variable, $V(\beta,\gamma)= V(\beta)$, and by using the wavefunction of each state as the multiplication of two one parametric functions as:

$$\Psi(\beta,\gamma,\Omega) = F_{L,\alpha}(\beta)\eta(\gamma,\Omega)$$

The Schrödinger equation can be separated into two equations:

$$\left[-\frac{1}{\beta^4}\frac{\partial}{\partial\beta}\beta^4\frac{\partial}{\partial\beta} + \frac{1}{4\beta^2}\left(4L(L+1) - 3\alpha^2\right) + u(\beta)\right]F_{L,\alpha}(\beta) = \varepsilon_\beta F_{L,\alpha}(\beta) \quad , \tag{2a}$$

$$\left[-\frac{1}{\langle\beta^2\rangle\sin 3\gamma}\frac{\partial}{\partial\gamma}\sin 3\gamma\frac{\partial}{\partial\gamma} + u(\gamma)\right]\eta(\gamma) = \varepsilon_\gamma\eta(\gamma) \quad , \tag{2b}$$

$L$ is the angular momentum and $\alpha$ presents its projection on the body-fixed $\hat{x}'$-axis and must be an even integer. Also, $\langle\beta_2\rangle$ is the average of $\beta_2$ over $F(\beta)$. The $\gamma$-angular part of this Hamiltonian is not the subject of our study on the spherical nuclei, but as presented in Ref. Bès [17], the eigenvalues of this part are given in terms of the seniority quantum number $v$ [10] and corresponding to the second-order SO(5) Casimir operator. Also, the details about the scheme of angular momenta realized in a fixed $v$ configuration are available in Ref. [5]. Now, if one applies the $f(\beta) = \beta^2 F(\beta)$ change on the $\beta$-part of the wavefunction, the new form of this part of Hamiltonian in a canonical-like form is:

$$\left[-\frac{\partial^2}{\partial\beta^2} + \frac{(v+1)(v+2)}{\beta^2} + v(\beta)\right]f(\beta) = \varepsilon f(\beta) \quad , \tag{3}$$

which $v(\beta) = (2B/\hbar^2)V(\beta)$ and $\varepsilon = (2B/\hbar^2)E(\beta)$ are the reduced potential and energy. Now, if we consider the potential in β shape variable as a harmonic oscillator with an energy-dependent string constant, $v(\beta,\varepsilon) = k(\varepsilon)\beta^2$, the energy of the considered systems with the dependence to only radial parameter is [5]:

$$\varepsilon = \sqrt{k(\varepsilon)}\left(N + \frac{5}{2}\right), \qquad (4)$$

In this formula, $N = 2n_\beta + v$ which $n_\beta$ is the number of oscillator quanta and $v$ is the seniority number. Budaca in Ref.[5], has used the most straightforward dependence of energy as, $k(\varepsilon) = 1 + a\varepsilon$, which the $a$ is the slope parameter and varies between $a = 0$ and $a = 1$ values which these limits correspond with $\beta$ vibration and $\beta$ rigidity. He showed that by this choice, the energy eigenvalues, Eq.(4), have a quadratic form and the positive solution which labeled by N is:

$$\varepsilon_N = \left[\left(N + \frac{5}{2}\right)\frac{a}{2} + \sqrt{1 + \left(N + \frac{5}{2}\right)^2 \frac{a^2}{4}}\right]\left(N + \frac{5}{2}\right), \qquad (5)$$

This new form of energy-dependent uses a part of the energy expended by the collective excitations to stabilize the nuclear shape against the $\beta$ oscillations and therefore, make it possible to describe the collective nuclear motion in a more realistic condition. This means, if the energy increased due to the $\beta$ oscillation, this term is stiffening the nucleus in its original spherical shape. To consider the effect of this stiffness parameter on the energy spectra, we considered some non-real systems in which their numbers of oscillator quanta are $N = 25, 50$ and $100$ and determined all of its $2^+$ energy levels in such systems which have different $a$ values. These levels are unfolded and analyzed by using the RMT, similar to experimental energy levels.

3. Method of analysis

The relation between the statistical properties of energy spectra and quantum chaos can be described by Random matrix theory. The work of Brody et al. suggests that the fluctuation properties of chaotic systems are well described by the GOE limit of RMT [29] which this hypothesis was verified by Gomez et al. [30] and also showed that, quantum chaos can be defined in terms of matrix theory. On the other hand, Berry and Robnik have demonstrated in their work [37], that Poisson distribution is in agreement with the distribution of energy levels fluctuations in the integrable systems. In this study, we have used the nearest neighbor spacing distribution (NNSD) to compare the statistical situation of considered sequences in comparison with both Poisson distribution ( to describe systems that have regular dynamics) [21]:

$$P(s) = e^{-s}, \qquad (6)$$

and the Gaussian orthogonal ensemble (GOE) limit of RMT which is given by the Wigner distribution [21]:

$$P(s) = \frac{1}{2}\pi s e^{-\frac{\pi s^2}{4}}, \qquad (7)$$

which exhibits the chaotic properties of spectra. The requirement of a complete and pure level scheme for analyses in the framework of NNSD, together with the lack of enough experimental data, forced us to combine levels with the same spin-parity assignments to construct such sequences with at least 25 members and make an exact analysis. In the first stage, one must separate the fluctuation and the smoothed average parts of the considered spectrum which the behavior of later is non-universal and cannot be described by RMT [29-30]. This procedure is known as unfolding and can be done via different formalisms which their results may yield different but report the same tendency comparing of different sequences. We used the method is commonly used in the statistical analyses of experimental data. The number of the levels below the *E* value is taken as [23]:

$$N(E) = \int_0^E \rho(E)dE = e^{(\frac{E-E_0}{T})} - e^{-\frac{E_0}{T}} + N_0 \quad , \tag{8}$$

$N_0$ describes the level number with energies less than zero and must assume zero. Now, we fit the *N(E)* (≡*F(E)*) function to provide a correct set of energies as [23]:

$$E_i' = E_{\min} + \frac{F(E_i) - F(E_{\min})}{F(E_{\max}) - F(E_{\min})}(E_{\max} - E_{\min}) \quad , \tag{9}$$

which makes both $E_{\max}$ and $E_{\min}$ unchanged under such conversion and also make a level density with the constant average. These $\{E_i'\}$ construct an unfolded sequence that is dimensionless and their average spacing is constant and equal to 1 but the actual spacing of these sequences exhibits strong fluctuation. In the considered sequences, the $s_i = (E'_{i+1}) - (E'_i)$ describe the nearest neighbor spacing between the members of sequences. Therefore, the *P(s)* distribution defines the probability for the $s_i$ to lie within the infinitesimal interval [*s*, *s+ds*]. A comparison between the histogram of NNS distribution in different sequences with both Poisson and GOE curves describes the statistical situation of considered systems. The intermediate position between the regular and correlated limits of the statistical properties of the majority of physical systems suggests the usage of such distributions which can describe both limits by variation of some quantities [29-40]. The Berry- Robnik distribution [37] is one of the popular distributions:

$$P(s,q) = [q + \frac{1}{2}\pi(1-q)s] \times \exp(-qs - \frac{1}{4}\pi(1-q)s^2) \quad , \tag{10}$$

which is derived by assuming the energy level spectrum and describes Poisson and Wigner limits with $q = 1$ and 0, respectively. The standard procedure to get the statistical situation of the considered sequence can sort as: i) comparison between the histogram of sequence with Berry- Robnik distribution and ii) extract its parameter via estimation techniques. We extended the Maximum likelihood (ML) technique to avoid the disadvantages of estimation methods such as the least square fitting (LSF) technique. The process of MLE estimation is presented in detail in Refs. [38,47,52-53]. Here, we outline the basic ansatz and summarize the results. First, we construct the likelihood function as a product of all $P(s)$ functions:

$$L(q) = \prod_{i=1}^n P(s_i) = \prod_{i=1}^n [q + \frac{1}{2}\pi(1-q)s_i] e^{-qs_i - \frac{1}{4}\pi(1-q)s_i^2} \quad , \tag{11}$$

then, by maximizing the likelihood function, the desired estimator yield as:

$$f : \sum \frac{1-\frac{\pi s_i}{2}}{q+\frac{\pi}{2}(1-q)s_i} - \sum (s_i - \frac{\pi s_i^2}{4}) \quad , \tag{12}$$

Now, we solve Eq.(12) by the Newton-Raphson method to get "$q$" with high accuracy[38]:

$$q_{new} = q_{old} - \frac{F(q_{old})}{F'(q_{old})} \quad .$$

The final result yield as:

$$q_{new} = q_{old} - \frac{\sum \frac{1-\frac{\pi s_i}{2}}{q_{old}+\frac{\pi s_i}{2}(1-q_{old})} + \sum s_i + \frac{\pi s_i^2}{4}}{\sum \frac{-(1-\frac{\pi s_i}{2})^2}{(q_{old}+\frac{1}{2}\pi(1-q_{old})s_i)^2}} \quad , \tag{13}$$

In this procedure, we use the prediction of LSF method as the initial value and the ML result corresponds to the converging values of iterations Eq.(13).

4. Results

In this paper, we analyzed the statistical fluctuation of different levels of even-even nuclei to identify possible correlation due to the vibration versus rigidity related to radial parameter. To this aim and in the first section of analyses, we used the latest available empirical data and also the theoretical prediction for the considered levels via a parameter-free method, e.g. Eq.(5) in such nuclei which are located in the $90 \leq A \leq 140$ mass region, have underlying U(5) symmetry and their $R_{4/2}$ energy ratio are in the 2.00 -2.15 region. Both experimental values and theoretical results are used to prepare different sequences which are categorized as their spin-parity assignment and quantum numbers which are used to label states in this model, e,g., $\beta$ oscillator quanta and seniority numbers. These sequences are unfolded as explained in Section3 and then compared with Berry-Robnik distribution. We extracted the "q" values via ML technique which estimate with more accuracy. Also, the analyses on the short sequences make an overestimation of chaoticity degrees, i.e., $q$. Therefore, we test a comparison between the amounts of this quantity in additional sequences and would not concentrate only on the implicit values of these quantities.

i) Classification by the spin-parity and different quantum numbers

In the first part, we test the effect of the different spins of considered levels on the statistical properties of these even-even nuclei. To this aim, both experimental values and theoretical predictions for $0^+$, $2^+$ and $4^+$ levels unfolded and then compared with Berry-Robnik distribution which its parameter extracted via Eq.(13). The results for the q values together the number of levels in each mass region are presented in Table 1 and Figure 1.

Table 1. The chaocity parameters of different levels by using both experimental and theoretical values, respectively, $q_{Exp.}$ and $q_{Th.}$. $N$ is the number of spacing of each spin.

| Sequence | N | $q_{Exp.}$ | $q_{Th.}$ |
|---|---|---|---|
| All levels | 188 | 0.39±0.08 | 0.37±0.05 |
| $0^+$ | 66 | 0.43±0.10 | 0.42±0.08 |
| $2^+$ | 75 | 0.33±0.07 | 0.35±0.05 |
| $4^+$ | 47 | 0.29±0.05 | 0.33±0.06 |

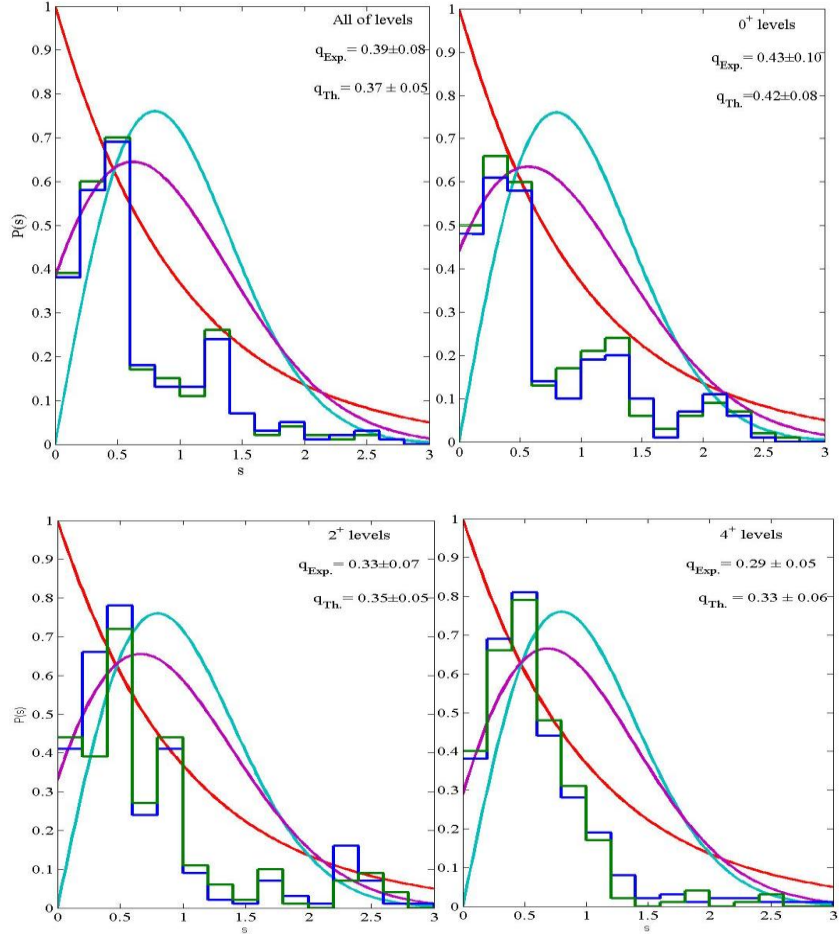

Figure 1. NNSD histograms of considered sequences in Table 1. Blue and green histograms are, respectively distributions of experimental values and theoretical predictions for considered levels. Also, red, turquoise and purple curves are respectively, Poisson, GOE and the Berry-Robnik distribution plotted by using the $q_{Exp.}$ values.

The results of these table and figure, show a correlated behavior in all of the considered sequences but when we consider levels with bigger spins, $4^+$, the maximum correlation is observed. This result is in agreement with our previous study [47] which suggests more GOE-like behavior when the spin of the considered levels increased. Also, the similarity between the q values which are determined by using both theoretical predictions and experimental values confirm the accuracy of this parameter-free technique in the exact description of energy levels in this mass and energy regions and therefore, we can extend the similar investigation for systems with total oscillator number up to N =100 which we could consider in the following.

In this part, all of the considered levels with different spins are labelled in the framework of the five-dimensional stiffening spherical vibrator model (SSV) [5] by using the number of $\beta$ oscillator quanta and seniority number as:

$N = 2n_\beta + v$ $\rightarrow$ $N = 5$ $\rightarrow$ $n_\beta = 0, 1$ and $2$

$v = 0, 1, 2, 3, 4$ and $5$

which for our considered levels and nuclei, the maximum value of seniority number is $v = 4$. Now, the new sequences are prepared by using experimental and theoretical values of these energies which are classified as their similar $n_\beta$ and $v$ values and their statistical fluctuations are compared to both regular and chaotic limits. The results are presented in Figure 2.

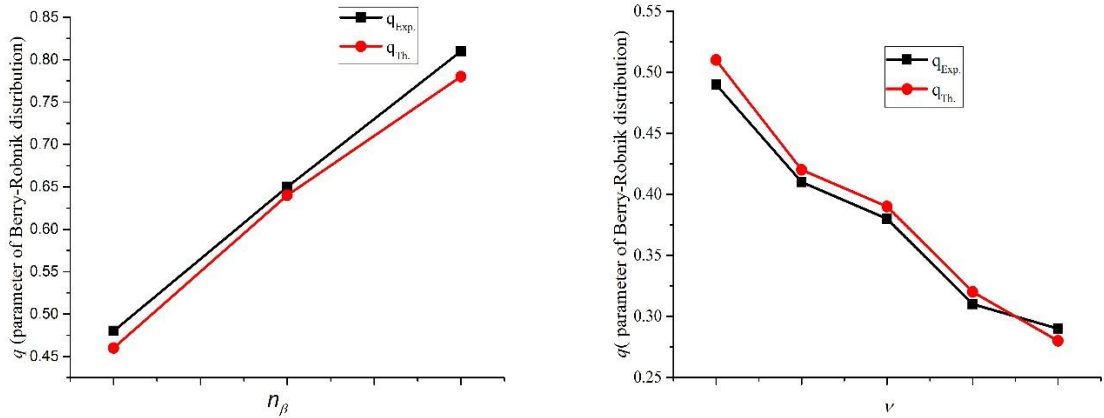

Figure 2. Variation of q values, chaocity measure, in relation to $\beta$ oscillator quanta and seniority using experimental and theoretical values.

The results show apparent dependence between the correlation of these sequences and the amounts of the considered quantum numbers. The maximum correlation is suggested for sequences that have the minimum oscillator quanta but maximum seniority number. The lack of enough experimental data makes it impossible to consider the relation between spin, seniority and $\beta$ oscillator quanta simultaneously for each spin separately. Due to this restriction, all of 188 levels are analyzed together which the results are presented in Figure 3.

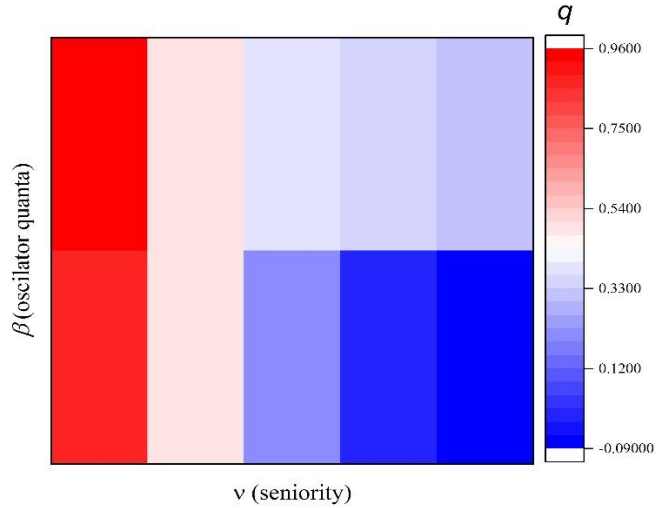

Figure3. chaocity measure (by using experimental energy values) in relation to both seniority and $\beta$ oscillator quanta for all of the 188 considered levels.

The results of these Table and Figures can be summarized as the follows:

1. A correlated behavior is observed for the considered nuclei which have underlying U(5) symmetry.

2. The predictions of the parameter-free five-dimensional stiffening spherical vibrator model for different energy levels are in good agreement with their experimental counterparts and suggest the same statistics in various categories. Also, this result confirms our assigned quantum numbers to different energy levels.

3. The observed correlated statistics for the levels that have the maximum seniority numbers may relate to the dominant SO(5) symmetry in such levels in comparison with the SO(3) symmetry.

4. if we relate the amount of $\beta$ oscillator quanta to the strength of vibrations of considered systems, the observed deviation from GOE limit in sequences with bigger $n_\beta$, suggests that the combination of different modes in energy spectra decreases correlation which we consider it in the following subsection.

ii) vibration versus rigidity

To compare the effect of rigidity on the spectral statistics of different systems, we considered the energy spectra, which are calculated by using the five-dimensional stiffening spherical vibrator model, Eq.(5). The results of the previous section confirm the accuracy of such formalism and the correctness of procedure which used to assign different quantum numbers of considered states. In this way and to make meaningful analyses, we considered systems with total harmonic oscillator quanta, N = 25, 50 and 100, similar to what has been done in Refs.[52-53]. In such systems, the number of $2^+$ levels are equal to respectively, 628, 1438 and 2226. The energy of these $2^+$ levels are determined by using Eq.(5) which the slope parameter, $a$, varied for them between $a = 0$ and $a = 1$ by step length $\Delta a = 0.1$. As have expressed in Ref.[5], these two $a = 0$ &1 limits are correspond respectively, with the dominant effect of $\beta$-vibration and $\beta$-rigidity on the spectral statistics. He showed that the spectrum would compress by using the negative values of *a* but when the positive values of this slope parameter were inserted in Eq.(5), the energy spectra would expand. Also, he pointed out *a* = 0 case corresponds entirely with the U(5) model with an energy ratio as $R_{4_1^+/2_1^+} \approx 2.00$. This means, the collective phenomena in the energy spectra are associated with only *a* > 0 case. These levels are unfolded and then compared with Berry-Robnik distribution whose parameter are extracted by the MLE method, Eq.(13) . The results are presented in Figure 4.

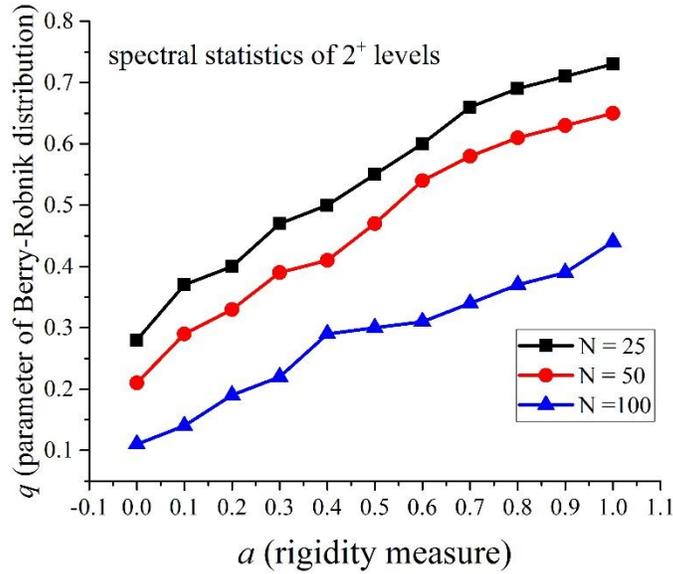

Figure 4. Spectral fluctuation of 2+ levels in different theoretical systems in relation to the strength of rigidity.

The results of this part of the analyses can be reported as the follows:

i) The maximum correlation is observed for $a = 0$ in all of the considered systems. The means, $\beta$ vibrations, make a correlated statistic which N = 100 system is located in the nearest position to GOE limit in comparison to other systems.

ii) when the rigidity increased in the considered systems, e.g. a →1, a deviation from correlated behavior is apparent in the considered systems which this Poisson-like behavior increases in the N=25 system in comparison to other systems. Paar et al. have suggested that, the rotation of nuclei contributes to the suppression of their chaotic dynamics which is known as the AbulMagd-Weidenmuller chaoticity effect [39]. Therefore, one may conclude that, rigidity decreases the impact of rotation and, consequently decreases the possible correlation of energy levels in such spherical nuclei. In the following study, we will consider the rigidity effect on the energy spectra of well-deformed nuclei, which are known as the candidates for SU(3) dynamical symmetry of IBM, by using a similar manner to make an exact conclusion about the effect of different modes on statistical fluctuations.

## 5. Summary and conclusion

Spectral statistics of even-even nuclei which have underlying U(5) symmetry are investigated by using both experimental data and theoretical predictions in the framework of the parameter-free five-dimensional stiffening spherical vibrator model. Various criteria are used to classify these levels in different sequences to make a sound relation between possible correlation and spin, seniority, the number of $β$ oscillator quanta and the rigidity strength. The Berry-Robink distribution is used to compare the statistical situation of these negative parity states in comparison with both Poisson and GOE limits. The apparent correspondence between the chaocity measure for the same categories which was prepared by using experimental data and theoretical predictions, approves the idea of using such a model in the investigation of energy spectra in the considered spherical nuclei. Also, the observed GOE-like behavior for systems that have the maximum seniority values, suggests the dominant impact of SO(5) symmetry in comparison with SO(3) on the correlation of the considered levels. The effect of rigidity is considered on the energy spectra of such systems with total oscillator quant equal to N =25, 50 and 100, which a deviation from correlated behavior is observed in the energy spectra when $β$- rigidity has its maximum values. We test the same analyses by using experimental data for other deformed and odd mass nuclei as the next part of this study to reach a general summary.

**Acknowledgement**

This work is published as a part of research project supported by the University of Tabriz Research Affairs Office.

**Author contributions**

H. Sabri and A. Hosseinnezhad performed the initial calculations, analyzed and interpreted the results, and wrote the main manuscript text. All authors commented on and reviewed the manuscript.

**Competing interests;**

The authors declare no competing interests.

**Data Availability Statement**

The datasets used and analyzed during the current study available from the corresponding author on reasonable request.